\documentclass[%
 reprint,superscriptaddress, amsmath,amssymb,
 aps, prd, nofootinbib
]{revtex4-2}
\usepackage{orcidlink}
\usepackage{graphicx}
\usepackage{dcolumn}
\usepackage{bm}
\usepackage[mathlines]{lineno}
\usepackage{array}
\usepackage[usenames,dvipsnames]{xcolor}
\usepackage[normalem]{ulem}
\usepackage{booktabs}
\usepackage{float}

\begin{document}

\preprint{APS/123-QED}

\title{Rapid Equation of State Inference from Multiprobe Datasets}
\author{Tamal RoyChowdhury\,\orcidlink{0000-0003-4242-0532}
}
\email{tamal@uwm.edu} \affiliation{University of Wisconsin-Milwaukee, Milwaukee, WI 53201, USA}
\author{Anarya Ray\,\orcidlink{0000-0002-7322-4748
}}
\email{anarya.ray@northwestern.edu} \affiliation{Center for Interdisciplinary Exploration and Research in Astrophysics (CIERA), Northwestern University, 1800 Sherman Ave,
Evanston, IL 60201, USA}
\affiliation{NSF-Simons AI Institute for the Sky (SkAI), 172 E. Chestnut Street, Chicago, IL 60611, USA}
\author{Jolien Creighton\,\orcidlink{0000-0003-3600-2406}}
\affiliation{University of Wisconsin-Milwaukee, Milwaukee, WI 53201, USA}
\author{Shaon Ghosh\,\orcidlink{0000-0001-9901-6253}}
\affiliation{Montclair State University, 1 Normal Avenue, Montclair, New Jersey 07043, USA}

\date{\today}

\begin{abstract}
Fully constraining the neutron star equation of state requires combining observations from various channels that probe different density regimes. We perform a joint multiprobe inference of the equation of state using gravitational-wave observations from binary neutron star mergers, NICER mass-radius measurements of X-ray pulsars, and radio timing mass measurements of massive pulsars. We constrain the parametrized equation of state and perform Bayesian model selection among a set of candidate equations of state. The joint analysis yields a radius of $R_{1.4} = 11.93^{+1.02}_{-0.83} \, \rm km$ for a canonical $1.4 \, M_\odot$ neutron star, and a maximum non-rotating neutron star mass of $M_{\rm max} = 2.31^{+0.23}_{-0.23} \, M_\odot$ with $90\%$ credible intervals. Gravitational-wave and NICER observations constrain the equation of state at low to intermediate densities, while massive pulsar observations constrain it at high densities by requiring the equation of state to support the observed neutron star masses. We find no simple correlation between the stiffness of an equation of state and Bayesian evidence across different observational probes.
\end{abstract}

\maketitle


\section{\label{sec:intro}Introduction}

The interior of a neutron star (NS) contains some of the densest forms of matter known to exist in the observable universe that are not reproducible in a laboratory environment~\citep{Lattimer:2006xb,Ozel:2016oaf}. The Tolman-Oppenheimer-Volkoff (TOV) equation~\citep{Oppenheimer:1939ne} describes the hydrostatic equilibrium between the gravitational force and the neutron degeneracy pressure inside an NS. In order to solve the TOV equation, we need an equation of state (EoS) that relates the degeneracy pressure to the density of matter in NSs. However, the EoS at supranuclear densities remains poorly constrained. The EoS of cold matter at extreme densities is expected to be universal \citep{Ozel:2016oaf,Lattimer:2012nd}; hence, all NSs are expected to share the same EoS. 

An EoS uniquely determines the mass-radius relationship of NSs~\citep{1992ApJ...398..569L}\footnote{Phase transitions can make $r_\varepsilon(m)$ multi-valued~\citep{Alford:2013aca,PhysRev.172.1325,Glendenning:1998ag}. In this work, we assume there is no phase transition in the EoS at high densities.}: 
\begin{equation}
p = p_\varepsilon(\rho) \Leftrightarrow r = r_\varepsilon(m),
\end{equation}
as well as their maximum mass. Hence, probes on NS structure such as measurements of the mass and/or radius, can be used to constrain the state of matter inside an NS. Observations of electromagnetic signals from X-ray pulsars and X-ray binaries~\citep{Choudhury:2024xbk,Miller:2021qha,Vinciguerra:2023qxq,Miller:2026vpr,Miller:2019cac,Riley:2019yda,Zhang:2006xz,2008MNRAS.387..268G,Bootsma:2025kns}, pulsar timing measurements of NSs in binary orbits~\citep{Ozel:2016oaf,Demorest:2010bx,Antoniadis:2013pzd,NANOGrav:2019jur,Damour:2009vw}, and binary neutron star (BNS) detections by LIGO-Virgo-KAGRA collaboration~\citep{LIGOScientific:2017vwq,LIGOScientific:2018hze,LIGOScientific:2018cki,LIGOScientific:2020aai,Flanagan:2007ix} provide complementary constraints to NS masses, radii, and tidal deformabilities. Due to the universality of the state of matter inside an NS, we can combine independent observations across multiple  channels to infer the common EoS and explore the properties of cold matter at extreme densities. Since the observed NSs are drawn from an underlying astrophysical population, the inferred EoS can also depend on the NS mass distribution. This has motivated hierarchical analyses that jointly constrain the EoS and NS population when combining a large number of observations~\citep{Wysocki:2020myz,Golomb:2021tll,Landry:2021hvl,Ghosh:2024cwc,Golomb:2024lds,Anik:2025vui,Ebiri:2026rwn,Koehn:2026nin}.

Several studies have measured the EoS relying on gravitational-wave~(GW) observations, either alone or in combination with X-ray mass-radius measurements, radio pulsar masses, electromagnetic counterparts to BNS mergers, and nuclear theory information~\citep{Biswas:2021pvm,Tang:2020koz,Tang:2021snt,Imam:2025lut,Biswas:2025tjt,Biswas:2025ivu,Jacobi:2026opf,Capano:2019eae,Ayriyan:2021prr,Essick:2020flb,Al-Mamun:2020vzu,Raaijmakers:2019dks,Dietrich:2020efo,Landry:2020vaw,Pang:2021jta,Ng:2025wdj,Rutherford:2024srk,Raaijmakers:2021uju, Imam:2024gfh, Imam:2024xqg}. Within this framework of Bayesian hierarchical inference, the EoS is constrained through two complementary approaches. Tabulated EoSs from theoretical descriptions of nuclear physics provide a discrete set of pressure-density relationships, enabling Bayesian model selection to assess which microphysical descriptions are better supported by astrophysical observations, albeit with greater dependence on the chosen models. On the other hand, parametrized EoS models such as the spectral decomposition of the adiabatic index~\citep{Lindblom:2010bb} and piecewise-polytropic parametrization~\citep{Read:2008iy} span a continuous space of possible pressure-density curves, allowing the data to constrain the EoS with fewer assumptions about the underlying physics. Additionally, nonparametric approaches place flexible priors over families of EoS functions, for example through Gaussian processes, allowing a broader range of pressure-density behavior without imposing a specific analytic form~\citep{Essick:2020flb,Landry:2018prl,Essick:2019ldf,Legred:2021hdx,Legred:2022pyp}. Together, these approaches provide complementary information: parametrized and non-parametrized inference characterize the allowed mass-radius and pressure-density relations, while model comparison determines how well specific microscopic descriptions are supported within those observational constraints. 

Combining multiprobe observations leverages the complementary constraining power of different probes; individual probes can be substantially offset from the joint result. A systematic study of which probes are sensitive to particular microphysical prescriptions for degenerate matter and to specific regions of the pressure-density plane is necessary to characterize biases in single-probe measurements and determine which observations would most effectively inform particular investigations in dense matter theory if additional observations were obtained. However, such a study requires repeated inference from combinations of single-channel datasets. Imposing an EoS prior during event level Bayesian parameter estimation can be computationally expensive, particularly as the number of observations grows. 
An efficient alternative is to reuse posterior samples from EoS-agnostic analyses and impose EoS priors through hierarchical inference and likelihood approximations. Several methods have been developed along these lines, particularly for GW observations~\citep{Ghosh:2021eqv,Golomb:2021tll,HernandezVivanco:2019vvk,HernandezVivanco:2020cyp,Wysocki:2020myz,Wouters:2025zju,Wouters:2026nbh}. 

In particular, the likelihood approximation scheme implemented in \texttt{GWXtreme}~\citep{Ghosh:2021eqv} reconstructs the relevant likelihood from existing posterior samples using density estimation, reduces its dimensionality by fixing tightly constrained source parameters and marginalizing over uninformative ones, enables rapid Bayesian model comparison comparison of candidate EoS models and empirical constraints on the pressure-density plane given GW data~\citep{Ghosh:2021eqv, Ray:2022hzg}. Its extension to hierarchical inference of parametrized EoSs~\citep{Ray:2022hzg} enabled rapid empirical constraints on the pressure-density plane using GW datasets. Extending both to multiprobe datasets and investigating the mentioned systematics of EoS inference across different probes 
is a necessary next step.

In this work, we extend the capabilities of \texttt{GWXtreme} to incorporate additional observation channels beyond GW data. We include NS mass-radius measurements of X-ray pulsars from NICER~\citep{2016SPIE.9905E..1HG} and precise mass measurements from radio timing of massive pulsars, and use these jointly with GW observations to perform parametrized EoS inference and Bayesian model selection among candidate EoS models. This enables us to compare how the relative support for candidate EoSs changes across individual observation channels and when the datasets are combined. We further identify which observations are more informative in different regions of the pressure-density and mass-radius planes, and obtain joint multiprobe constraints on the NS EoS; we show that restricting to individual probes can substantially affect the inferred conclusions. 

The paper is organized as follows. Section~\ref{sec:methods} describes the methodology behind the likelihood approximation scheme and Bayesian model selection, followed by the joint inference of parametrized EoS. In Sec.~\ref{sec:result} we show the results obtained by applying the multiprobe \texttt{GWXtreme} to the observed data and discuss the implications of these results. In Sec.~\ref{sec:disc} we summarize our key results, highlight caveats, and discuss future work.

\section{\label{sec:methods}Methods}

We constrain the NS EoS by combining independent observations from three different probes: GW detections of BNS mergers by the LVK, X-ray observations of pulsar hotspots by NICER, and mass measurements from radio observations of heavy pulsars. The data from these three channels are denoted by $\vec{d}_\alpha$ and each source is described by its own set of parameters denoted by $\vec{\theta}_\alpha$, where $\alpha$ denotes the probe, i.e. $\alpha \in \{\rm{GW,\, X,\, R}\}$ for GWs from BNS, NICER pulsars, and radio observations of heavy pulsars, respectively. Individual source parameters measured using Bayesian methods can provide information on EoS models that impose deterministic relationships between these parameters (see Sec.~\ref{sec:apdx}). Combining independent observations, one can therefore compute the Bayesian evidence of an EoS model given our multiprobe dataset as follows:

\begin{equation}
    p(\vec{d}_{\rm GW},\,\vec{d}_{\rm X},\,\vec{d}_{\rm R}|\mathcal{E}) = \prod_{\alpha} \prod_{i=1}^{N_\alpha}p(d_{\alpha,i}|\mathcal{E}),  
    \label{eq:joint_likelihood}
\end{equation}
where $N_\alpha$ is the total number of observations in the channel $\alpha$
\begin{equation}
    p(d_{\alpha,i}|\mathcal{E}) = \int_{\vec{\theta}_{\alpha, i}} p(d_{\alpha,i} \, | \, \vec{\theta}_{\alpha, i})\,p(\vec{\theta}_{\alpha, i} \, | \, \mathcal{E},pop)\,d\vec{\theta}_{\alpha, i}
\end{equation}
is the evidence for the EoS model given the data from a single observation, and $\vec{\theta}_{\alpha, i}$ are the EoS-sensitive parameters of the $i^{\rm th}$ observation in the channel $\alpha$. The term $p(\vec{\theta}_{\alpha, i} \, | \, \mathcal{E},pop)$ is the hierarchical prior in the parameters $\vec{\theta}_{\alpha, i}$ informed jointly by an EoS model $\mathcal{E}$ and the assumed NS population model ($pop$). The deterministic EoS prior restricts the integrand to a lower-dimensional subspace, preventing Monte Carlo summation over the EoS-agnostic posterior samples. The likelihood $p(d_{\alpha,i} \, | \, \vec{\theta}_{\alpha, i})$ as an analytic function of EoS-sensitive parameters $\vec{\theta}_{\alpha, i}$ marginalized over the other observables is usually intractable. Instead, we density estimate posterior samples from an EoS-agnostic parameter estimation analysis and use Bayes' theorem, integrating the likelihood numerically on the lower-dimensional region that has non-zero support from the EoS. 
A more detailed derivation and validation of the above method is presented in Refs.\citep{Ghosh:2021eqv, Ray:2022hzg}. We arrive at the following expression for the evidence in favor of an EoS model:

\begin{equation}
\begin{aligned}
&p(\vec{d}_{\rm GW},\,\vec{d}_{\rm X},\,\vec{d}_{\rm R}|\mathcal{E})  = \prod_{\alpha} \prod_{i=1}^{N_\alpha}
         \int_{S_{\mathcal{E}}}
         \frac{\hat{p}(\vec{\theta}_{\alpha,i}|d_{\alpha,i})}
              {\pi_{PE}(\vec{\theta}_{\alpha,i})} \\
&\qquad\qquad\qquad\qquad\qquad\qquad\qquad \times p(\vec{\theta}_{\alpha,i}|\mathcal{E},pop)
         \,d\vec{\theta}_{\alpha,i}
\end{aligned}
\end{equation}
where $\hat{p}(\vec{\theta}_{\alpha,i}|d_{\alpha,i})$ is the kernel density estimate of the posterior of the EoS-sensitive parameters, $\pi_{PE}(\vec{\theta}_{\alpha,i})$ is the prior used to obtain the EoS-agnostic posterior samples,  $p(\vec{\theta}_{\alpha,i}|\mathcal{E},pop)$ is the prior informed by the EoS and the astrophysical population distribution of the sources, and $S_{\mathcal{E}}$ is the lower-dimensional hypersurface of the parameter space where there is non-zero support for the EoS model. In the current work, we restrict ourselves to only the EoS-informed priors and leave joint population fits for future work, since biases due to misspecified population priors are expected to be smaller than measurement uncertainty for less than $\mathcal{O}(10)$ observations~\citep{Wysocki:2020myz,Ebiri:2026rwn}. 

For a tabulated set of theoretical EoS models, we compare how well each model describes the multiprobe dataset by computing the Bayes factor between pairs of models:

\begin{equation}
    \mathcal{B} = \frac{P(\vec{d}_{\rm GW},\,\vec{d}_{\rm X},\,\vec{d}_{\rm R}|\mathcal{E}_1)}{P(\vec{d}_{\rm GW},\,\vec{d}_{\rm X},\,\vec{d}_{\rm R}|\mathcal{E}_2)}
\end{equation}
where $\mathcal{E}_{1}$ and $\mathcal{E}_{2}$ denote two known EoS models. Alternatively, one can phenomenologically parametrize the EoS pressure-density relation, for example, through spectral or piecewise-polytropic parametrizations. We can compute the posterior distribution of these EoS \textit{hyperparameters}~$(\vec{\gamma})$ from our multiprobe likelihood function and prior knowledge of NS physics:

\begin{align}
\begin{split}
    p(\vec{\gamma}|\vec{d}_{\rm GW},\,\vec{d}_{\rm X},\,\vec{d}_{\rm R}) \propto p(\vec{\gamma}, I) \prod_{\alpha}& \prod_{i=1}^{N_\alpha} \int_{S_{\mathcal{E}}(\vec{\gamma})}\frac{\hat{p}(\vec{\theta}_{\alpha,i}|d_{\alpha,i})}{\pi_{PE}(\vec{\theta}_{\alpha,i})} d\vec{\theta}_{\alpha,i}
    \label{eq:evidence_gamma}
\end{split}
\end{align}
where $p(\vec{\gamma},I)$ is the hyperparameter prior, informed by our prior knowledge of NS physics denoted by $I$. The choice of EoS priors is discussed in  Sec.~\ref{sec:eos_priors}. The posterior distribution in Eq.~(\ref{eq:evidence_gamma}) is sampled stochastically using Markov chain Monte Carlo (MCMC) ensemble sampling. For each posterior sample $\vec{\gamma}_j$, we evaluate the pressure $p(\rho,\vec{\gamma})$ on a grid of density values $\rho$. Thus, at each density, we obtain $N_\mathrm{samples}$ pressure values, which form the posterior pressure distribution at that density. We compute the $5^{\rm th}$ and $95^{\rm th}$ percentiles of this distribution at each density and use them to construct the $90\%$ credible region of the inferred pressure-density relation. For details of parametrized models and priors, see Appendix~\ref{sec:apdx}.

\section{\label{sec:result}Results}

We have combined datasets from three different observation channels to infer the EoS. For GW sources, we use EoS-agnostic posterior samples for mass and tidal parameters (defined in Sec.~\ref{sec:apdx}) of GW170817~\citep{LIGOScientific:2017vwq,LIGOScientific:2018hze,LIGOScientific:2018cki} and GW190425~\citep{LIGOScientific:2020aai} from the LVK release. For the NICER sources, we use the mass and radius samples from the data release of pulsars J0437-4715~\citep{Choudhury:2024xbk}, J0740+6620~\citep{Miller:2021qha}, J0030+0451~\citep{Vinciguerra:2023qxq}, and J0614+3329~\citep{Miller:2026vpr}. For the radio pulsars, we use the mass measurements for the three heavy pulsars J1614–2230~\citep{Demorest:2010bx}, J0348+0432~\citep{Antoniadis:2013pzd}, and J0740+6620~\citep{NANOGrav:2019jur}. Information about the relevant parameters of each observation that we consider is listed in Tables~\ref{tab:GW} - \ref{tab:Radio}. 

\begin{table}[H]
    \centering
    \begin{tabular}{l c c c}
        \midrule
        BNS Events &  $\mathcal{M} \, (M_\odot)$ & $q$ & $\tilde{\Lambda}$\\
        \midrule 
        & & &\\[-0.9em]
         GW170817 & $1.186^{+0.001}_{-0.001}$ & $(0.73,1)$ & $300^{+500}_{-190}$\\[0.2em]
         GW190425 & $1.44^{+0.02}_{-0.02}$ & $(0.8,1)$ & $\leq 600$\\[0.2em]
         \hline
    \end{tabular}
    \caption{GW observations used in this analysis. The table lists the BNS parameters obtained from the corresponding analyses~\citep{LIGOScientific:2017vwq,LIGOScientific:2018hze,LIGOScientific:2018cki,LIGOScientific:2020aai}.}
    \label{tab:GW}
\end{table}

\begin{table}[H]
    \centering
    \begin{tabular}{l c c}
        \midrule
        PSR &  $\rm{Mass} \, (M_\odot)$ & $\rm{Radius} \, (\rm{km})$ \\
        \midrule
        & & \\[-0.9em]
         J0437–4715 & $1.418^{+0.004}_{-0.004}$ & $11.36^{+0.95}_{-0.63}$ \\[0.2em]
         J0740+6620 & $2.08^{+0.07}_{-0.07}$ & $13.7^{+2.6}_{-1.5}$ \\[0.2em]
         J0030+0451 & $1.34^{+0.16}_{-0.15}$ & $12.71^{+1.27}_{-1.18}$ \\[0.2em]
         J0614+3329 & $1.44^{+0.07}_{-0.07}$ & $11.16^{+1.61}_{-1.27}$ \\[0.2em]
         \hline
    \end{tabular}
    \caption{X-ray pulsars observed by NICER~\citep{2016SPIE.9905E..1HG} used in the analysis. The table lists the NSs with the corresponding mass and radius measurements~\citep{Choudhury:2024xbk,Miller:2021qha,Vinciguerra:2023qxq,Miller:2026vpr}. }
    \label{tab:NICER}
\end{table}

\begin{table}[H]
    \centering
    \begin{tabular}{l c}
        \midrule
        PSR &  $\rm{Mass} \, (M_\odot)$ \\
        \midrule 
        & \\[-0.9em]
         J1614–2230 & $1.928^{+0.017}_{-0.017}$ \\[0.2em]
         J0348+0432 & $2.01^{+0.04}_{-0.04}$ \\[0.2em]
         J0740+6620 & $2.14^{+0.10}_{-0.09}$ \\[0.2em]
         \hline
    \end{tabular}
    \caption{Heavy radio pulsars and corresponding NS masses used in this work~\citep{Demorest:2010bx,Antoniadis:2013pzd,NANOGrav:2019jur}. }
    \label{tab:Radio}
\end{table}

\begin{figure*}[t!]
    \centering
    \includegraphics[width=0.9\linewidth]{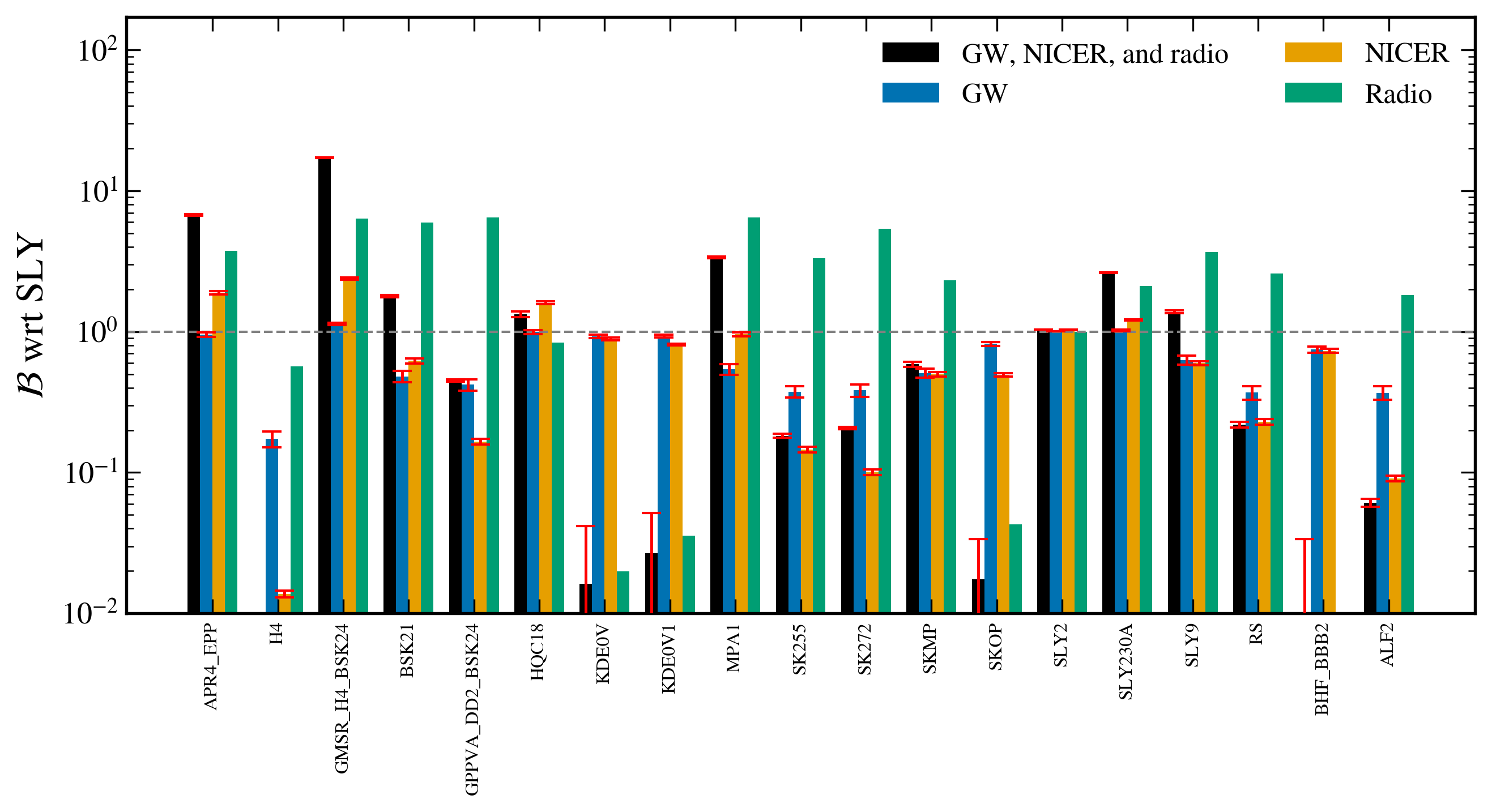}
    \caption{Bayes factors for a set of EoS models relative to SLY obtained using joint GW, NICER, and radio datasets, and individual datasets. The dashed horizontal line at $\mathcal{B} = 1$ denotes equal evidence relative to SLY, while values above (below) unity denote greater (lower) support than SLY. Red error bars show twice the standard deviation of Bayes factors obtained using the bootstrap approach for GW and NICER posterior samples, with corresponding uncertainties propagated to joint analysis. No error bars are shown for the radio-only Bayes factors because the likelihood is calculated directly from the mass measurements rather than from a KDE reconstruction. Differences in the Bayes factors obtained from GW, NICER, and radio observations indicate the complementary sensitivity of the different observations to the EoS. See Tab.~\ref{tab:eos_bayes_factors} for the Bayes factors.}
    \label{fig:joint_bf_SLY}
\end{figure*}

Given our multiprobe dataset, we first compare a tabulated set of known EoS models, each of which corresponds to a specific description of cold nuclear matter. In Fig.~\ref{fig:joint_bf_SLY}, we show the Bayes factors (relative to SLY~\citep{Chabanat:1997un,Douchin:2001sv,Haensel:2004nu}, which is a moderately soft EoS that predicts small tidal deformability and weak tidal effects) evaluated using individual GW, NICER, and radio datasets, as well as their joint combination. For the GW and NICER datasets, we show uncertainties in the Bayes factors arising from the finite number of posterior samples used to construct the KDE likelihood. These uncertainties are estimated by bootstrap resampling the posterior samples~\citep{Ghosh:2021eqv}, reconstructing the KDE for each realization, and then repeating the model selection calculation. The error bars in Fig.~\ref{fig:joint_bf_SLY} correspond to twice the standard deviation of the resulting Bayes factor distribution. Since the radio likelihood is evaluated directly from the measured mass and not using a KDE, we do not assign such an uncertainty to the radio Bayes factors.

The Bayes factor comparison reveals preferences among the different observation channels. The GW and NICER measurements provide comparable support for a number of EoSs relative to SLY, implying that several models remain consistent with the mass-radius and tidal-deformability constraints probed by these observations. However, radio measurements produce stronger differences between the models, preferring EoS models that can provide sufficient pressure at high densities to support massive NSs.

The joint analysis reflects the complementary sensitivity of these observations. Among the models chosen, GMSR\_H4\_BSK24~\citep{Grams:2022lci,Pearson:2018tkr} receives the strongest support from the joint dataset. This EoS combines the GMSR(H4) description of the inner crust and core~\citep{Grams:2022lci} with the BSk24 prescription for the outer crust~\citep{Pearson:2018tkr}. APR4\_EPP~\citep{Read:2008iy,Akmal:1998cf}, BSK21~\citep{Goriely:2010bm}, MPA1~\citep{Muther:1987xaa}, and several SLY-family models receive greater support than SLY. In contrast, models such as H4~\citep{Lackey:2005tk}, KDE0V, KDE0V1~\citep{Agrawal:2005ix}, and SKOP~\citep{Reinhard:1995zz} are disfavored by the joint observations. Importantly, the Bayes factor ranking does not correspond to a simple soft-to-stiff ordering of the EoS models, indicating that the observations are sensitive to the detailed density dependence of the pressure rather than a single overall measure of stiffness. The uncertainty estimates do not change the overall ranking of the strongly favored or disfavored models, but they become relevant when interpreting Bayes factors between otherwise comparable EoS models.

\begin{figure}
    \centering
    \includegraphics[width=0.48\textwidth]{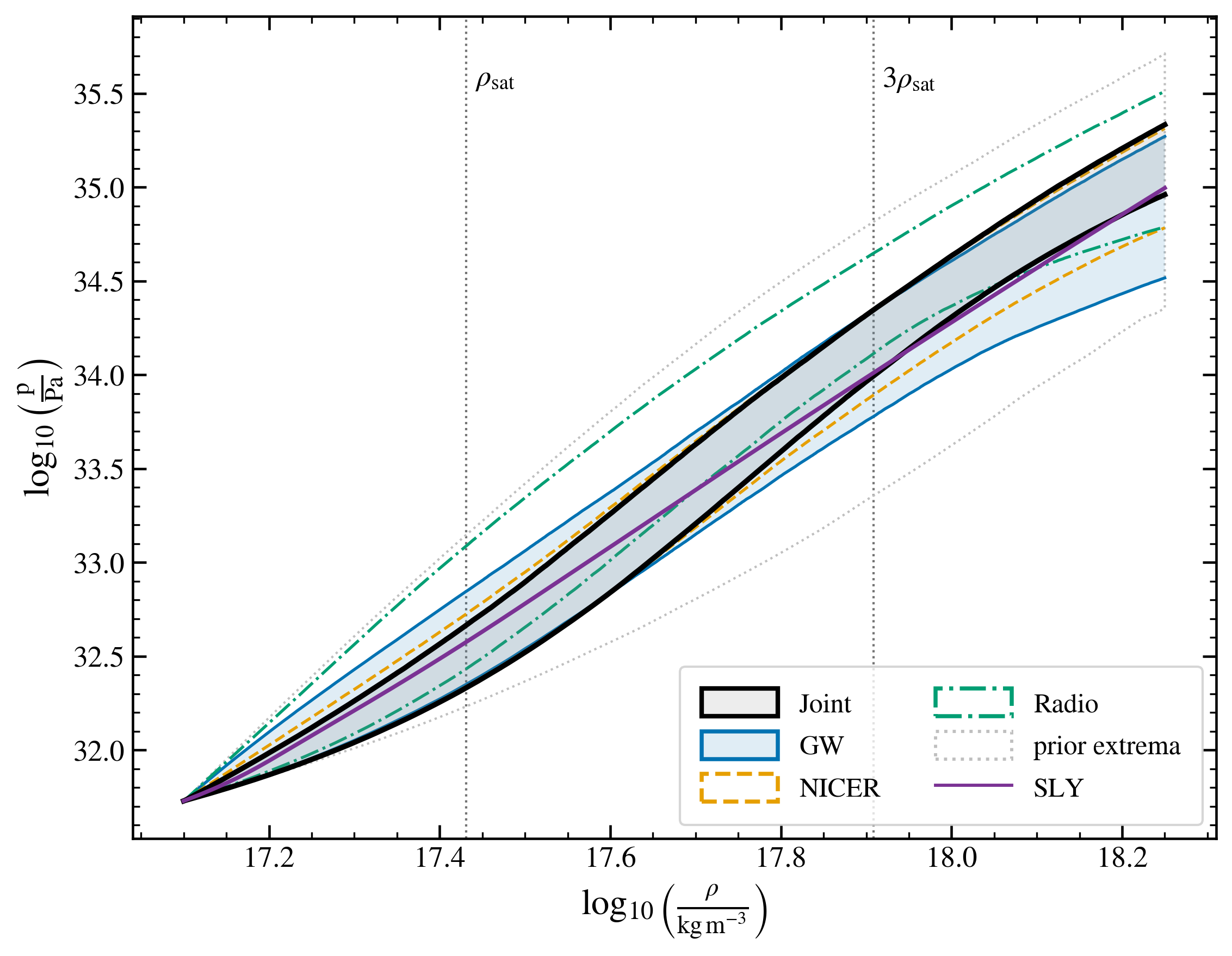}
    \caption{Constraints on the posterior distribution of EoS in the pressure-density plane, obtained using GW, NICER, and radio observations. The plots show $90\%$ credible bounds of the inferred pressure as a function of density for joint and individual datasets, denoted by the shaded regions. The dashed gray lines denote the EoS prior extrema. The joint inference provides tighter constraints on the pressure-density relation by combining information from observations that are sensitive to different density regimes.}
    \label{fig:p_vs_rho}
\end{figure}

The differences between the individual and joint Bayes factors illustrate the role of multiprobe observations. \textbf{The joint inference contains substantially more information, while analyses based on individual probes can yield conclusions that are significantly offset from the joint result.} Some EoS models, such as GPPVA\_DD2\_BSK24~\citep{Typel:2009sy,Grill:2014aea,Pearson:2018tkr}, receive relatively strong support from radio observations, but weaker support from  GW and NICER observations, suggesting greater consistency with the high-density constraints from massive pulsars than with the intermediate-density radius and tidal constraints from GW and NICER observations. Similar behavior is seen for models like SK272~\citep{Agrawal:2003xb} and RS~\citep{Friedrich:1986zza}, which are strongly supported by radio observations but disfavored by the GW and NICER datasets. The joint model selection therefore preferentially supports EoS models that simultaneously reproduce the properties of a canonical NS and provide sufficient pressure support at the higher densities relevant to massive pulsars. Thus, strong support from a single observation channel does not necessarily result in strong joint evidence.

\begin{figure}
    \centering
    \includegraphics[width=0.48\textwidth]{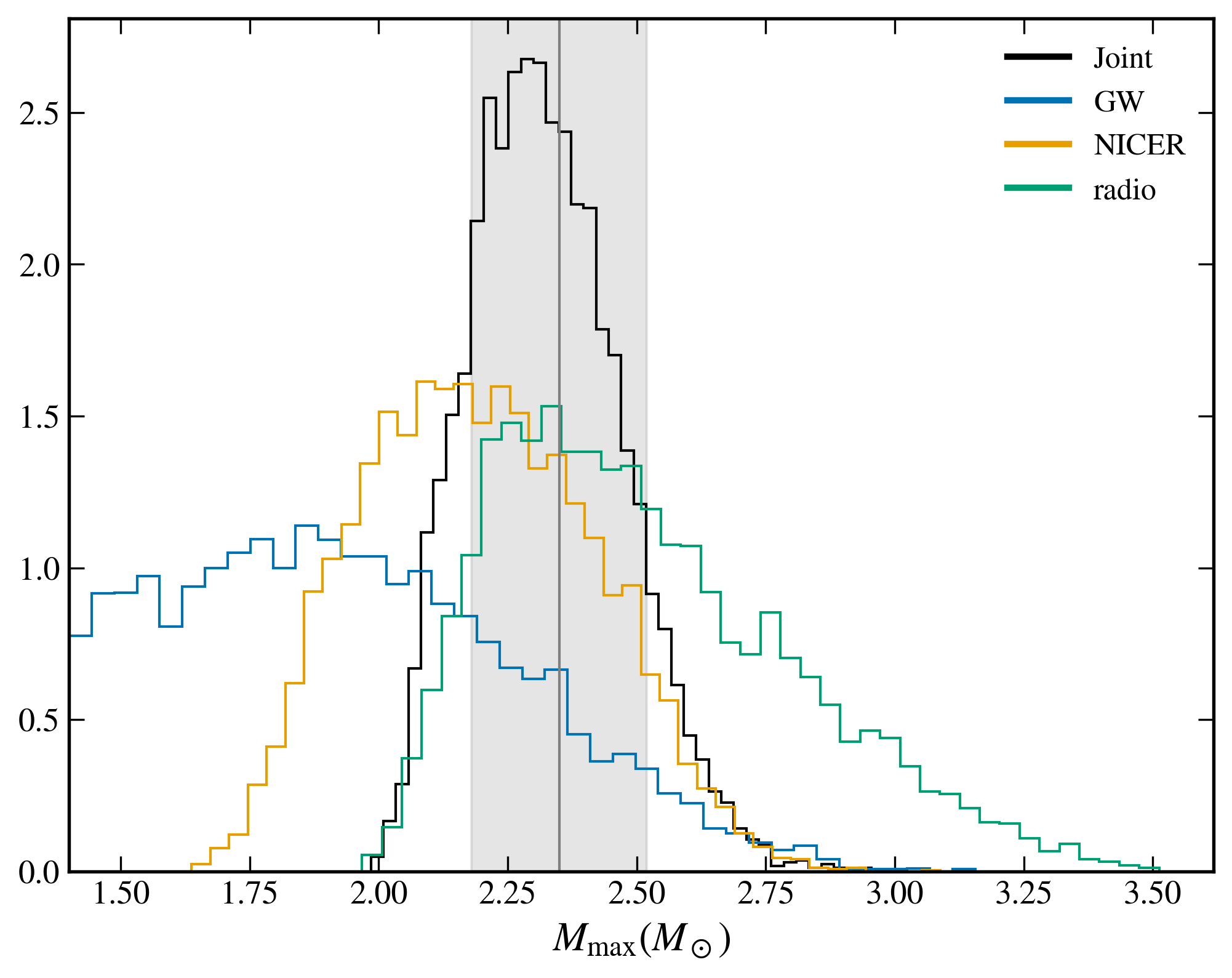}
    \caption{Posterior distribution of the maximum mass of a NS $M_{\rm{max}}$ inferred using GW, NICER, and radio observations alone, as well as jointly. Radio observations of massive pulsars constrain the lower bound of $M_{\rm{max}}$. Combined inference yields the narrowest constraints. The gray vertical band indicates the mass of the heaviest measured galactic pulsar J0952-0607~\citep{Romani:2022jhd} at $2.35 \, \pm \, 0.17 \, M_{\odot} $.}
    \label{fig:m_max_14}
\end{figure}
\begin{figure}
    \centering
    \includegraphics[width=0.48\textwidth]{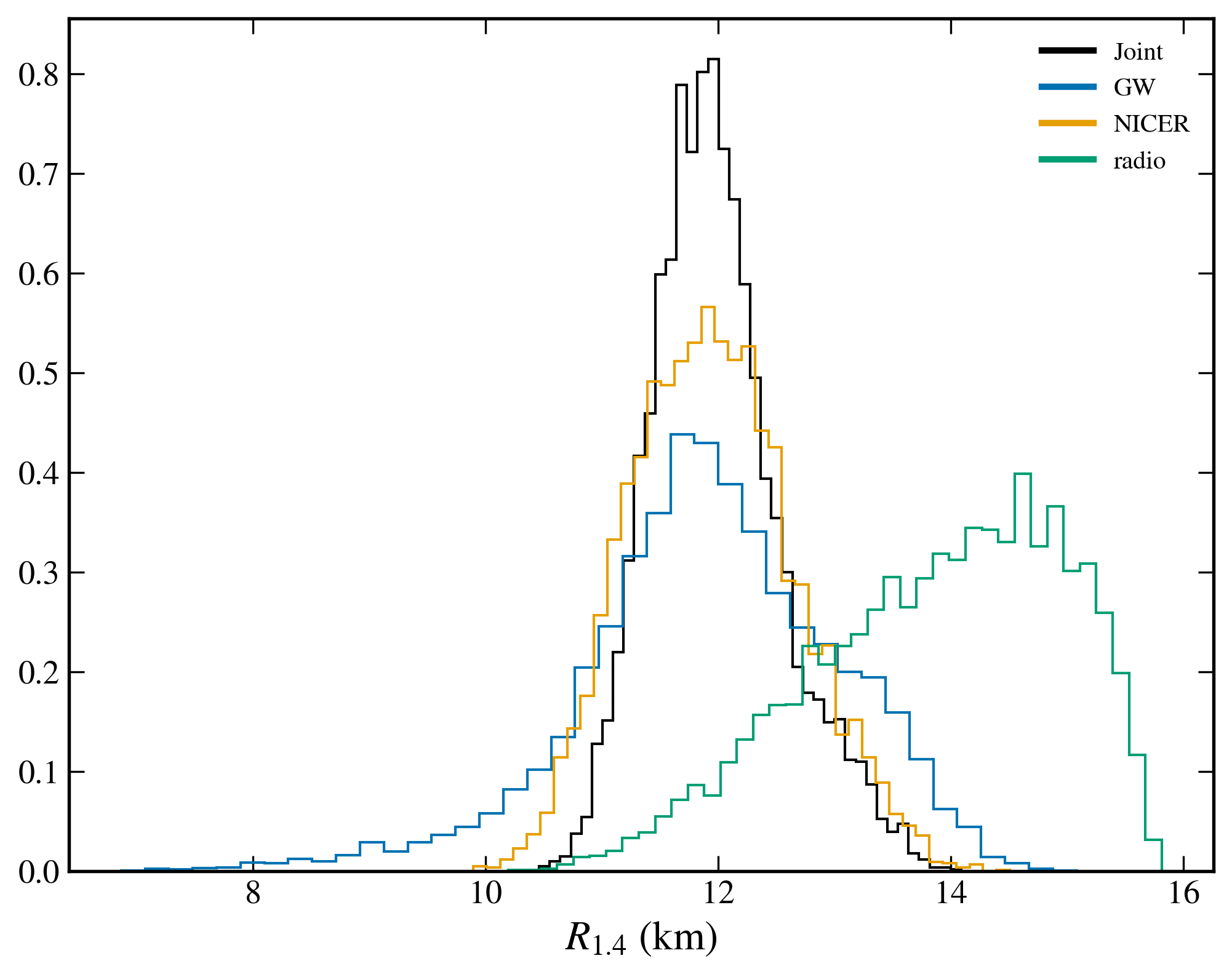}
    \caption{Posterior distribution of the radius of a $1.4M_\odot$ NS $R_{\rm{1.4}}$ inferred using GW, NICER and radio observations alone, as well as jointly. GW and NICER measurements provide the primary constraints on the NS radius.}
    \label{fig:r_14}
\end{figure}

To further investigate the necessity of multiprobe inference and the impact of restricting to single probes, we now turn to parametric pressure-density constraints. Figure~\ref{fig:p_vs_rho} shows the EoS constraints obtained by running \texttt{GWXtreme} for various combinations of datasets. \textbf{Each individual probe constrains complementary regions of the pressure-density plane;  
the joint constraint is substantially narrower.} The red dashed vertical line in Fig.~\ref{fig:p_vs_rho} denotes the nuclear saturation density $\rho_{\rm{sat}}$. \textbf{In the intermediate-density regime, approximately ($\rho_{\rm sat}\lesssim \rho \lesssim 3\rho_{\rm sat}$), both GW and NICER observations constrain the pressure relative to the prior}, reflecting their sensitivity to NS radii and tidal deformabilities. \textbf{At higher densities ($\rho \gtrsim 3\rho_{\rm sat}$), the radio pulsar measurements become more informative}, requiring the EoS to provide sufficient pressure support for the most massive observed NSs. The complementary sensitivity of the individual probes is also reflected in the posterior distributions of the maximum non-rotating neutron-star mass, $M_{\rm max}$, and the canonical radius, $R_{1.4}$, shown in Figures~\ref{fig:m_max_14} and~\ref{fig:r_14}, respectively.

\begin{figure*}[t!]
    \centering
    \includegraphics[width=0.9\textwidth]{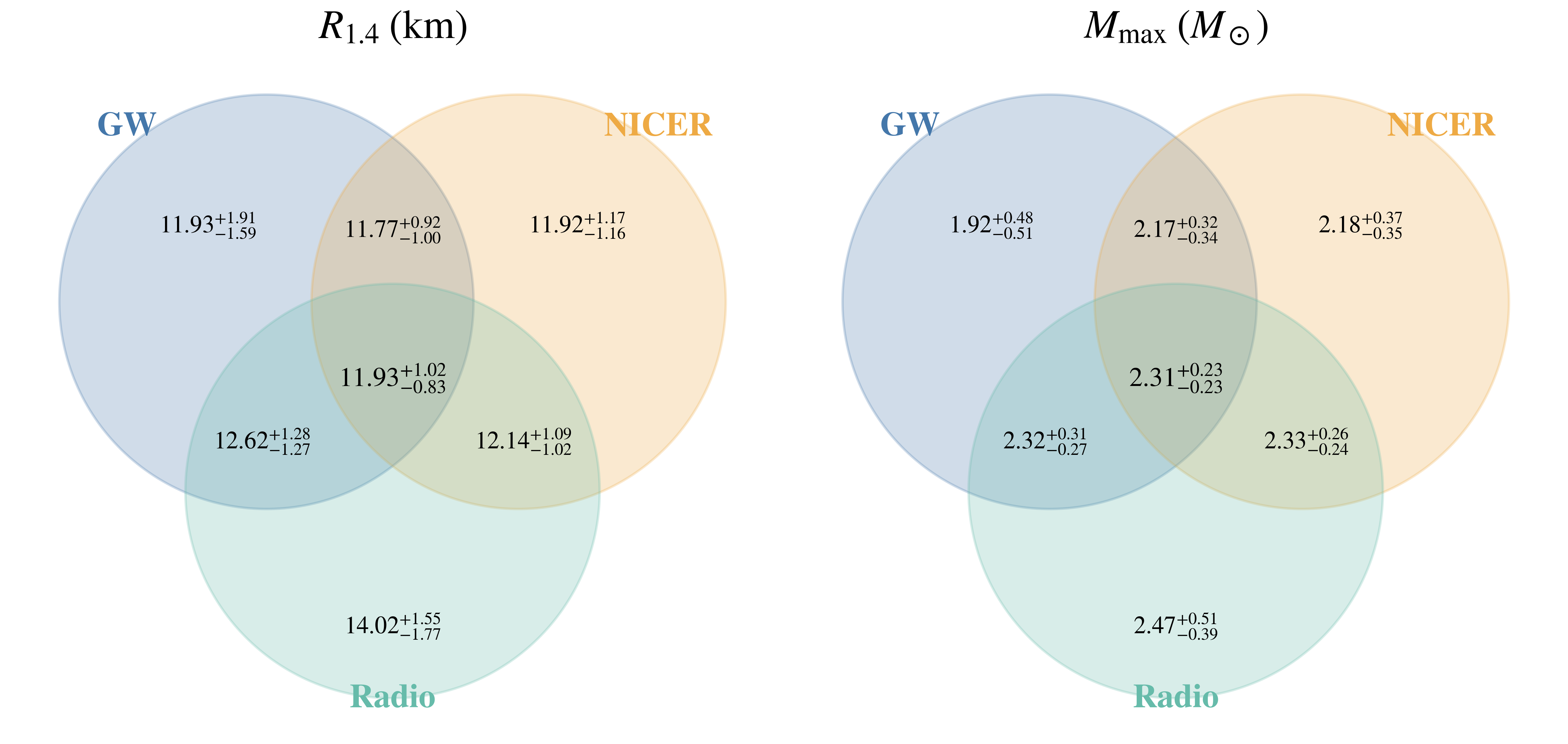}
    \caption{Comparison of the inferred radius of a canonical ($1.4\,M_\odot$) NS on the left, and the maximum mass of a non-rotating NS on the right, inferred using GW, NICER and radio observations obtained individually and in combinations. Each dataset combination shows the median of the posterior along with the $90\%$ credible intervals. The $R_{1.4}$ constraints from GW and NICER observations are broadly consistent, yielding a joint value close to $ 12 \, \rm km$, while radio observations prefer a larger radius and have less impact on the joint constraints. However, radio observations have a larger impact on $M_{\rm max}$, shifting the inferred maximum mass to a higher value in the joint constraint. This plot highlights the complementary sensitivity of the different observation channels to the NS mass and radius. }
    \label{fig:venn_combined}
\end{figure*}

The $M_{\rm max}$ posteriors further illustrate the different roles of the individual probes. GW observations yield a broad distribution extending to relatively low maximum masses, while NICER provides a more constrained posterior. Although neither probe directly measures $M_{\rm max}$, their sensitivity to tidal deformability~\citep{Hinderer:2007mb,Flanagan:2007ix,Damour:2009vw,Binnington:2009bb} and radius restricts the range of allowed low- and intermediate-density pressure profiles and, within the EoS parametrization, indirectly suppresses realizations with very large maximum masses. In contrast, massive radio pulsars place a strong lower bound on $M_{\rm max}$ by requiring the EoS to support stars with masses of approximately $2\,M_\odot$, while providing comparatively little information on how far above the observed pulsar masses the maximum supported mass may lie. The joint analysis combines these effects to yield $M_{\rm max}=2.31^{+0.23}_{-0.23}\,M\odot$ at $90\%$ credibility. A similar complementarity is seen in $R_{1.4}$ posteriors. The GW and NICER posteriors overlap because both are sensitive to the structure of canonical-mass NSs, whereas the radio-only inference favors somewhat larger radii through correlations between high-density pressure support and the lower-density pressure that determines stellar radii~\citep{Lattimer:2000nx}.

\begin{figure}
    \centering
    \includegraphics[width=0.48\textwidth]{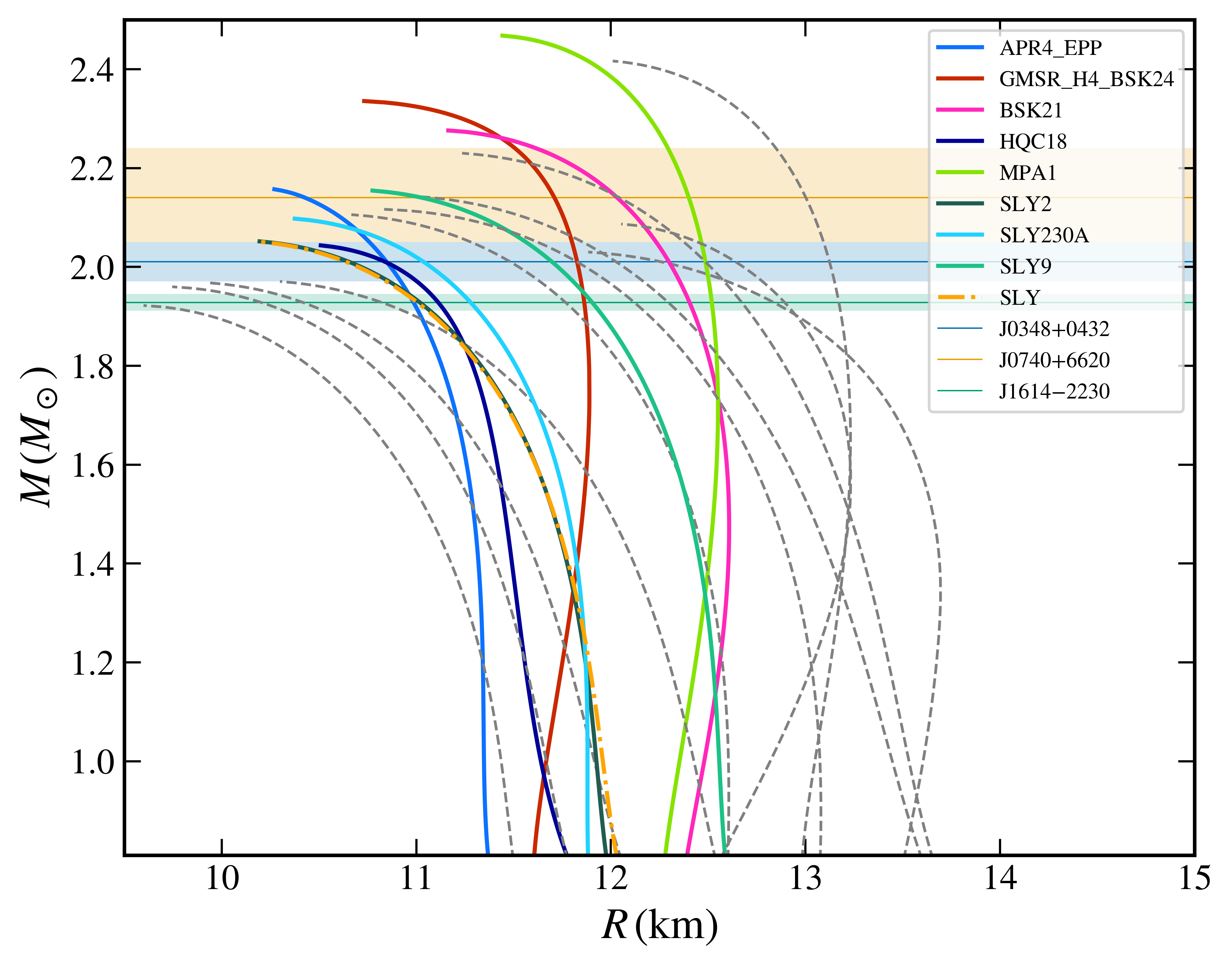}
    \caption{EoS curves in the $M$-$R$ space. The gray lines represent EoS that have a $\mathcal{B}<1$ relative to SLY as shown in Fig.~\ref{fig:joint_bf_SLY}. The colored curves represent EoS with $\mathcal{B}>1$, while the dashed yellow line is the curve for SLY. The Bayes factors are calculated using the combined GW, NICER X-ray pulsar, and radio pulsar observations. The colored bands represent the measured masses of the radio pulsars and their uncertainties. Support for an EoS is determined by a combination of tidal and radius constraints at low-to-intermediate densities and sufficient pressure support at high densities to support massive NSs.}
    \label{fig:mr_radio}
\end{figure}

Figure~\ref{fig:mr_radio} plots the mass-radius curves for various known EoS models. The colored curves denote models that are preferred over SLY according to the joint Bayes factors in Fig.~\ref{fig:joint_bf_SLY}, while the gray curves denote models that are disfavored. The preferred models span a broad range of mass-radius behavior, demonstrating that the ability to support ($\sim2\,M_\odot$) NSs alone is not sufficient to determine whether an EoS is favored, because several disfavored models also satisfy this requirement. Instead, the model preference reflects the ability of a given EoS to simultaneously reproduce the radius and tidal constraints at low to intermediate densities and the high-density pressure support required by massive NSs. This illustrates \textbf{the complementarity between the parametrized and model selection analyses}. The former constrains pressure-density behavior and macroscopic mass-radius relations, while the latter identifies which specific microphysical descriptions of dense matter better reproduce the multiprobe observations~\citep{Lattimer:2006xb,Burgio:2021vgk}.

\section{\label{sec:disc}Discussion}

We present a joint hierarchical inference of the NS EoS using a multiprobe dataset that combines GW events, X-ray pulsars, and radio pulsars. We perform Bayesian model selection among a set of candidate EoS models given multiprobe and single-probe datasets. The resulting Bayes factor ranking does not follow a simple stiffness ordering and is sensitive to the detailed density dependence of the pressure. The individual probes also provide information over different density regimes. Using a four-parameter spectral parametrization, we obtain joint constraints on the EoS. GW observations constrain the tidal response of NSs and provide sensitivity to low-to-intermediate densities, while NICER measurements constrain the mass-radius relationship and are primarily sensitive to intermediate densities. Radio measurements of massive pulsars constrain the high-density behavior by requiring the EoS to support the observed NS masses. The joint constraints favor EoSs that produce relatively compact canonical-mass NSs while retaining sufficient pressure support at high densities. We obtain $R_{1.4} = 11.93^{+1.02}_{-0.83} \, \rm km$ and $M_{\rm max} = 2.31^{+0.23}_{-0.23} \, M_\odot$, where the quoted values are the posterior medians with $90\%$ credible intervals. These results are consistent with previous multimessenger studies~\citep{Dietrich:2020efo,Landry:2020vaw,Pang:2021jta,Ng:2025wdj}, which combined BNS observations with subsets of NICER X-ray observations used in this work, as well as electromagnetic counterparts such as gamma-ray bursts and kilonovae. 

An advantage of the \texttt{GWXtreme} likelihood approximation framework is that it reuses posterior samples from EoS-agnostic analyses, allowing additional BNS observations to be incorporated without requiring full EoS-dependent parameter estimation runs for each model. This substantially reduces the cost of incorporating additional BNS observations expected from future observing runs. As the ground-based detector network approaches upgraded design sensitivities, the number of BNS detections is expected to increase substantially, further motivating efficient hierarchical inference and model-selection techniques~\citep{KAGRA:2013rdx,Magee:2022kkc}. However, additional developments are necessary to enable scalable, high-fidelity EoS inference with growing multiprobe datasets.

While we combine multiple observation channels in the current work, we do not account for the population properties of NSs in the different channels. With more observations expected during the O5 observing run, population information could become increasingly important~\citep{Wysocki:2020myz,Ebiri:2026rwn}. In addition, to represent GW and NICER posteriors, our current method relies on kernel density estimation (KDE). The computational cost of evaluating a KDE grows with the number of posterior samples, making this approach progressively less efficient as more events are included. A natural extension is therefore to replace the existing KDE with a more scalable density estimator such as a normalizing flow~\citep{2019arXiv191202762P,2019arXiv190809257K,Wouters:2025zju,Quinn-Vitabile:2026xyz,Thakur:2026smp,Thakur:2026dpr}. Additionally, the framework could be extended to jointly infer NS population parameters by simultaneously fitting population models and allowing them to differ for different observation channels. Lastly, our model uses the TOV solver in \texttt{LALSuite}~\citep{lalsuite}, which assumes a non-rotating NS. This approximation is generally valid for BNS systems in which the component spins are small~\citep{LIGOScientific:2018hze,Brown:2012qf,Miao:2021gmf}. However, this approximation is less accurate for NICER pulsars~\citep{Riley:2019yda,Miller:2021qha,Bogdanov:2021yip} since these pulsars have millisecond periods with non-negligible rotational effects and a TOV solver that accounts for rotation could therefore provide more accurate constraints. Accelerated high-fidelity solvers for axisymmetric NS structures~\citep{Ray:2026fti} are being integrated into GWXtreme to mitigate rotation-induced biases that may exceed measurement uncertainty for future X-ray detectors such as eXTP~\citep{Li:2025uaw}.

\section{\label{sec:ack}Acknowledgments}
The authors would like to thank Caleb Ogle and Joseph David Quinn-Vitabile for comments on the manuscript. The authors would like to thank Lami Suleiman for conducting the internal review. The authors are grateful for the computational resources provided by the LIGO Laboratory supported by National Science Foundation Grants No. PHY-0757058 and No. PHY-0823459. This work was supported by the National Science Foundation
Grants No. PHY-2513124. This material is based upon work supported by NSF's LIGO Laboratory which is a major facility fully funded by the National Science Foundation. The authors also acknowledge the use of Large Language Models (LLMs) in this work, specifically OpenAI's ChatGPT (GPT - 5.6 Sol), for plotting code, polishing text, and reviewing the literature.

This document has LIGO preprint number LIGO-P2600617.

\section{\label{sec:apdx}Appendix}

\subsection{\label{sec:GWs}GW signals from BNS mergers}

For GW signals from BNS mergers, the EoS-informative parameters are given by

\begin{align}
\begin{split}
    \vec{\theta}_{\rm GW} &= \{ m_1, m_2, \lambda_1, \lambda_2 \} \\
    \lambda_i &= \frac{2}{3G}k_2\,m_i\,r_i^5
    \label{eq:gw_param}
\end{split}
\end{align}
where $r_i$ $(i = 1,2)$ are the radii of the component NSs and $k_2$ is the tidal Love number~\citep{Hinderer:2007mb,Flanagan:2007ix,Damour:2009vw,Binnington:2009bb}. An EoS model parametrized using $\vec{\gamma}$ (see Sec.~\ref{sec:eos_priors}) gives the pressure-density and mass-tidal-deformability relations.

\begin{equation}
    p = p(\rho,\vec{\gamma}), \;\; \lambda = \lambda(m,\vec{\gamma})
    \label{eq:eos_relation}
\end{equation}
Hence, we get a marginalized hierarchical likelihood

\begin{align}
\begin{split}
    p(d_{\rm GW}|\vec{\gamma}) & = \int p(d_{\rm GW} |  m_1, m_2, \lambda_1, \lambda_2) \, p(\lambda_1,\lambda_2|m_1,m_2,\vec{\gamma}) \\ 
     & \qquad \times p(m_1, m_2)\, dm_1 \,dm_2\, d\lambda_1 \, d\lambda_2
     \label{eq:likelihood_GW_2}
\end{split}
\end{align}
The dimensionality of this integral can be reduced using the likelihood approximation scheme described in~\citep{Ghosh:2021eqv,Ray:2022hzg} by reparametrizing $(m_1,m_2,\lambda_1,\lambda_2)$ to chirp mass $(\mathcal{M})$, mass ratio $(q)$, and tidal parameters $(\tilde{\Lambda},\delta\tilde{\Lambda})$ that are defined as 

\begin{equation}
    \label{eq:chirp_mass}
    \mathcal{M} = \frac{(m_1m_2)^{3/5}}{(m_1+m_2)^{1/5}}
\end{equation}

\begin{equation}
    \label{eq:mass_ratio}
    q = \frac{m_2}{m_1}\,,\,(m_1>m_2)
\end{equation}

\begin{equation}
    \label{eq:lambda}
    \begin{aligned}
        \tilde{\Lambda} &= \frac{8}{13}\Bigl[ (1+7\eta-31\eta^2)(\Lambda_1+\Lambda_2) \\
        &\quad + \sqrt{1-4\eta}(1+9\eta-11\eta^2)(\Lambda_1-\Lambda_2)\Bigr]
    \end{aligned}
\end{equation}

\begin{equation}
    \label{eq:lambda_tilde}
    \begin{aligned}
        \delta \tilde{\Lambda} &= \frac{1}{2} \Bigl[ 
        \sqrt{1-4\eta} \left(1-\frac{13272}{1319}\eta + \frac{8944}{1319}\eta^2\right)  
        (\Lambda_1 + \Lambda_2) \\
        &\quad + \left( 1 - \frac{15910}{1319}\eta + \frac{32850}{1319} \eta^2 + \frac{3380}{1319}\eta^3   \right) (\Lambda_1-\Lambda_2)
        \Bigr]
    \end{aligned}
\end{equation}

In the above equations, the symmetric mass ratio $\eta$ and dimensionless tidal-deformability $\Lambda_i$ ($i=1,2$) are defined as

\begin{equation}
    \label{eq:eta}
    \eta = \frac{m_1m_2}{(m_1+m_2)^2}
\end{equation}

\begin{equation}
    \label{eq:lambda_nondim}
    \Lambda_i = \lambda_i G \left(\frac{c^2}{Gm}\right)^5
\end{equation}
where $\lambda_i$ is the tidal deformability defined in Eq.~\ref{eq:gw_param}. Note that we assume $m_1>m_2$ when defining Eqs.~\ref{eq:chirp_mass} - \ref{eq:lambda_nondim}. We use the fact that the single-event likelihood given the EoS-sensitive parameters multiplied by the EoS-independent priors on those parameters is proportional to the posterior distribution of those parameters given GW data. Thus, we have

\begin{align}
\begin{split}
    p(d_{\rm GW}|\vec{\gamma}) &= \int \frac{p({\mathcal{M},q,\tilde{\Lambda},\delta\tilde{\Lambda}|d_{\rm GW}})}{\pi_{\rm PE}({\mathcal{M},q,\tilde{\Lambda},\delta\tilde{\Lambda}})} \, p(\tilde{\Lambda},\delta\tilde{\Lambda}|\mathcal{M},q,\vec{\gamma}) \\
    & \qquad \times \, p(\mathcal{M},q)\,d\mathcal{M}\,dq\,d\tilde{\Lambda}\,d\delta\tilde{\Lambda}
\end{split}
\end{align}
where $\pi_{\rm PE}({\mathcal{M},q,\tilde{\Lambda},\delta\tilde{\Lambda}})$ is the uninformative prior used for EoS-agnostic PE runs. This prior can be simplified to be uniform in tidal deformabilities, i.e., $\pi_{\rm PE}({\mathcal{M},q,\tilde{\Lambda},\delta\tilde{\Lambda}}) \propto p(\mathcal{M},q)$ as discussed in~\citep{Ghosh:2021eqv,Ray:2022hzg}. Furthermore, chirp mass is well measured and posteriors are sharply peaked around the mean of the samples of $\mathcal{M}$. $\tilde{\Lambda}$ provides the dominant tidal contribution, while the posterior is comparatively insensitive to $\delta\tilde{\Lambda}$. Hence, the integral simplifies to a one-dimensional integral.

\begin{equation}
    \begin{split}
        p(d_{\rm GW}|\vec{\gamma}) & \propto \int p({\mathcal{M},q,\tilde{\Lambda},\delta\tilde{\Lambda}|d_{\rm GW}}) \\
        & \qquad \times \,p(\tilde{\Lambda},\delta\tilde{\Lambda}|\mathcal{M},q,\vec{\gamma}) \, d\mathcal{M}\,dq\,d\tilde{\Lambda}\,d\delta\tilde{\Lambda}  \\     
        &\propto \int p(q,\tilde{\Lambda}(\bar{\mathcal{M}},q,\vec{\gamma})|d_{\rm GW})\, dq
        \label{eq:evidence_gw}
    \end{split}
\end{equation}
To evaluate the integral numerically, we use EoS-agnostic PE samples and approximate them using a Gaussian KDE that is corrected for edge effects.

\subsection{\label{sec:Xrays}X-ray observations using NICER}
For NICER observations, the parameter samples are given by

\begin{equation}
    \vec{\theta}_{\rm X} = \{ m, r\}
    \label{eq:xray_param}
\end{equation}
A parametrized EoS model defines the mass-radius relation

\begin{equation}
    r = r(m,\vec{\gamma})
\end{equation}
Thus, the marginalized hierarchical likelihood becomes

\begin{align}
\begin{split}
    p(d_{\rm X}|\vec{\gamma}) & = \int p(d_{\rm X} | m,r) \, p(m) p(r|m,\vec{\gamma}) \,dm \,dr \\
    \label{eq:likelihood_xray}
\end{split}
\end{align}
As in the GW case, the posterior distribution of mass and radius is a product of the likelihood and the priors on mass and radius. Hence, we get

\begin{equation}
    \begin{split}
        p(d_{\rm X}|\vec{\gamma}) & = \int \frac{p(m,r|d_{\rm X})}{\pi_{\rm PE}(m,r)} \, p(m)p(r|m,\vec{\gamma}) \,dm \,dr \\
        &= \int \frac{p(m,r|d_{\rm X})}{\pi_{\rm PE}(m,r)} \, p(m)\delta(r-r(m,\vec{\gamma})) \,dm \,dr \\
        &= \int \frac{p(m,r(m,\vec{\gamma})|d_{\rm X})}{\pi_{\rm PE}(m,r(m,\vec{\gamma}))} \, p(m) \,dm \\
    \end{split}
\end{equation}
where the numerator can again be approximated by using a Gaussian KDE on the mass-radius samples. The integral is performed up to the maximum mass allowed by the EoS described by the hyperparameters $\vec{\gamma}$.

\subsection{\label{sec:radios}Radio observations of massive pulsars}

For the radio observations, we use the mass measurements with uncertainties. 

\begin{equation}
    \vec{\theta}_{\rm R} = \{ m \}
\end{equation}
For each pulsar, we approximate the measured mass posterior as a Gaussian distribution whose median and width are determined by the reported mass and credible interval, assuming a flat prior on the mass. For a given EoS specified by $\vec \gamma$, only NS masses below the corresponding maximum supported mass $M_{\rm max}(\vec \gamma)$ are allowed. Therefore, the marginalized likelihood is obtained by integrating the posterior up to the maximum allowed mass. 
\begin{equation}
    \begin{split}
        p(d_{\rm R}|\vec{\gamma}) = \int^{M_{\rm max}(\vec \gamma)}_0 p(m|d_{\rm R})\,dm
    \end{split}
\end{equation}
Thus, the radio likelihood measures the fraction of the observed pulsar mass posterior that is compatible with the maximum mass supported by a given EoS. 

\subsection{\label{sec:eos_priors}EoS priors}

We use a four-parameter spectral parametrization of the EoS that expands the adiabatic index in powers of the logarithm of pressure. This avoids the non-differentiability at the junctions between consecutive segments that occurs in piecewise-polytropic parametrization. Under the four-parameter spectral parametrization, we have

\begin{equation}
    \mathrm{ln} \, \Gamma(p,\vec{\gamma}) = \sum_{k=0}^3 \, \gamma_k\, \left( \mathrm{ln} \frac{p}{p_o} \right)^k
    \label{eq:spectral_param_eq}
\end{equation}
where $\Gamma(p)$ is the adiabatic index at pressure $p$ and $p_o$ is the minimum pressure above which this equation is valid. We can also compute the pressure as a function of the energy density using the thermodynamic relation

\begin{equation}
    \frac{de}{dp} = \frac{e+p}{p\, \Gamma(p,\vec{\gamma{}})}
    \label{eq:energy_pressure}
\end{equation}

An EoS characterized by a particular $\vec{\gamma}$ is considered physical if it is thermally stable, causal, and gives a maximum NS mass greater than the mass of any observed NS to date~\citep{Wysocki:2020myz,Carney:2018sdv}. Thermal stability requires $de/dp > 0$, which is satisfied by imposing $\Gamma(p,\vec{\gamma}) \in [0.6,4.5]\, \forall \, p \in (5.37 \times10^{32},1.19 \times 10^{38}) \, \mathrm{dyn \, cm^{-2}}$. Additionally, priors on $\vec{\gamma}$ are uniform: $\gamma_0 \in [0.2,2],\, \gamma_1 \in [-1.6,1.7], \gamma_2 \in [-0.6,0.6], \, \gamma_3 \in [-0.02,0.02]$. Causality requires that the speed of sound in NS matter, $c_s = \sqrt{dp/de}$, remain below the speed of light up to the central pressure $p_{c,\rm{max}}$ of the maximum-mass configuration. Note that we allow a $10\%$ numerical tolerance $(c_s<1.1c)$ to avoid rejecting otherwise viable EoS realizations because of interpolation and numerical differentiation errors near the causal boundary. The maximum mass constraint is satisfied by requiring $m_{\mathrm {max}} > 1.4 M_\odot$.  Combining all these priors gives

\begin{widetext}
\begin{equation}
    p(\vec{\gamma},I) \propto
    \begin{cases}
        1 & \text{where} \; 0.2\leq\gamma_0\leq2, \, -1.6\leq\gamma_1\leq1.7,\,-0.6\leq\gamma_2\leq0.6,\,0.02\leq\gamma_3\leq0.02, \\
        & \qquad  \;\;\; 0.6\leq\Gamma(p,\vec{\gamma})\leq4.5,\, c_s(p_{c,\mathrm{max}},\vec{\gamma})<1.1c, \, m_\mathrm{max}(\vec{\gamma})>1.4M_\odot \\
        0 & \text{otherwise}
    \end{cases}
    \label{eq:hyperparameter_priors}
\end{equation}
\end{widetext}


\begin{table*}[t]
\centering
\caption{Bayes factors for different EoS models relative to SLY, calculated for the three observation channels individually, and jointly.}
\label{tab:eos_bayes_factors}
\begin{tabular*}{\textwidth}{@{\extracolsep{\fill}}lcccc@{}}
\midrule
Equation of state & GW & NICER & Radio & Joint \\
\midrule
APR4\_EPP~\citep{Read:2008iy,Akmal:1998cf} & $0.954$ & $1.89$ & $3.75$ & $6.76$ \\
H4~\citep{Lackey:2005tk} & $0.173$ & $0.0137$ & $0.569$ & $0.00136$ \\
GMSR\_H4\_BSK24~\citep{Grams:2022lci,Pearson:2018tkr} & $1.13$ & $2.38$ & $6.35$ & $17.2$ \\
BSK21~\citep{Goriely:2010bm} & $0.482$ & $0.62$ & $5.98$ & $1.79$ \\
GPPVA\_DD2\_BSK24~\citep{Typel:2009sy,Grill:2014aea,Pearson:2018tkr} & $0.42$ & $0.166$ & $6.48$ & $0.451$ \\
HQC18~\citep{Baym:2017whm} & $0.996$ & $1.6$ & $0.836$ & $1.34$ \\
KDE0V~\citep{Agrawal:2005ix} & $0.925$ & $0.887$ & $0.0198$ & $0.0162$ \\
KDE0V1~\citep{Agrawal:2005ix} & $0.929$ & $0.811$ & $0.0355$ & $0.0267$ \\
MPA1~\citep{Muther:1987xaa} & $0.542$ & $0.96$ & $6.49$ & $3.38$ \\
SK255~\citep{Agrawal:2003xb} & $0.374$ & $0.146$ & $3.33$ & $0.182$ \\
SK272~\citep{Agrawal:2003xb} & $0.383$ & $0.101$ & $5.37$ & $0.207$ \\
SKMP~\citep{Gulminelli:2015csa,1989PhRvC..40.2834B,Danielewicz:2008cm} & $0.508$ & $0.498$ & $2.32$ & $0.588$ \\
SKOP\citep{Reinhard:1995zz} & $0.821$ & $0.494$ & $0.0429$ & $0.0174$ \\
SLY2~\citep{Chabanat:1997un} & $1.01$ & $1.03$ & $1$ & $1.04$ \\
SLY230A~\citep{Chabanat:1997un} & $1.02$ & $1.21$ & $2.12$ & $2.63$ \\
SLY9~\citep{Chabanat:1997un} & $0.631$ & $0.599$ & $3.67$ & $1.39$ \\
RS~\citep{Friedrich:1986zza} & $0.369$ & $0.229$ & $2.58$ & $0.218$ \\
BHF\_BBB2~\citep{Baldo:2014yja} & $0.747$ & $0.732$ & $0.000364$ & $0.000199$ \\
ALF2~\citep{Alford:2004pf} & $0.369$ & $0.0907$ & $1.82$ & $0.061$ \\
\hline
\end{tabular*}
\end{table*}

\nocite{*}

\bibliography{apssamp}

\end{document}